A simple statistical approach to prediction in open high dimensional chaotic systems

M. LuValle

Rutgers University

Abstract

Two recent papers on prediction of chaotic systems, one on multi-view embedding[1], and the second on prediction in projection[2] provide empirical evidence to support particular prediction methods for chaotic systems. Multi-view embedding[1] is a method of using several multivariate time series to come up with an improved embedding based predictor of a chaotic time series. Prediction in projection[2] discusses how much smaller embeddings can provide useful prediction even though they may not be able to resolve the dynamics of the system. Both papers invoke a nearest neighbor[3], or Lorenz method of Analogue (LMA)[4] approach to estimation. However with open high dimensional chaotic systems there may be no very close nearest neighbor trajectories in a history, so in this paper we add in the thread of linear response theory[5], although our approach is quite simple assuming that linear regressions[6] on multiple embeddings[5] will be sufficient. The approach is thus to create linear response models[5] of multiple[1] low dimensional embeddings[2] to provide practical methods of predicting in high dimensional, open chaotic systems. Some theory is developed, mostly around two unproven conjectures to suggest methods of prediction, and they are applied to prediction in the earth's climate system (predicting regional rainfall in a small region) multiple seasons ahead and to power a simulated automated trading program applied to three stock indexes, the Dow Jones industrial average, the Dow Jones transportation average, and the Dow Jones utility average, based on data from 1929 through 2007[7].

Introduction:

The focus of this paper is on two ideas, and their apparent empirical success. The first idea has to do with sampling delay maps from the set of possible delay maps. Rather than considering the set of all possible delay maps of a given size, we consider only disjoint partitions of a given set of delay maps. Not only does this reduce the computational effort in getting a reasonable sampling of the set of delay maps, it appears to improve the predictive power as well. The second idea is reframing a continuous prediction problem as a threshold passing problem. When this reframing is possible, a weakly predictive approach can become usefully predictive.

In the first example the statistical model is initially estimated on a climate model[8,9] simulation of 100 years at constant global surface air temperature (GSAT) created by setting greenhouse gases constant and letting GSAT stabilize. Since there are no disturbances being generated externally, the fluctuations being picked up are fluctuations generated by the unstable manifold, which is precisely what the linear response theory is concerned with[5]. Subsequent analysis using real data is simply to eliminate poor predicting regression models (possibly caused by departures of the computer model from real physics) and to counter the effect of the random predictor variable effect. The thresholding is not necessary for the 1st example.

For the Dow Jones averages, the history is used for estimating the linear response functions. In this case, as per Ruelle[5], it is expected that external disturbances to the attractor will dominate as a form of noise, and the unstable manifold regression will be significantly swamped. By reframing the prediction problem as identifying when the system will pass a threshold, removed from the mean, allows a variety of statistical tactics to be used successfully to take advantage of prediction. We experimented with shrinkage, stronger down selection on good predictors, and trading off between increasing the amount of data used in initial fitting and used in model down selection.

Both systems, are non-stationary. For climate, under the current consumption of fossil fuels, GSAT is increasing. Thus the thermal energy driving the movement of air and water are changing, and one expects the attractor to change with it. In high dimensional systems the transition is smooth[10,11]. Therefore approximating the dynamics with mixtures of attractors in a neighborhood of a given GSAT drawn from a climate model is reasonable assuming the climate model is itself a reasonable representation. For the stock market, the transitions are more abstract. For example, before 1929, speculation and manipulation of the stock market was legal. In the 1950s, the NYSE promoted "own a piece of american business"[12] diversifying and increasing the set of people trading stock on the market. The performance of a simple automated algorithm based on selling when the stock was predicted to drop and buying back when it doesn't, does very well after the 1950's, but not very well before, with progressively worse performance the less regulated the market becomes, and the less diverse the set of investors becomes going back in time.

In the next section some conjectures are stated and theory is developed based on those conjectures to aid in predicting trajectories in high dimensional systems using embedology and linear response. The third section discusses the results of applying these methods to two examples.

**Conjectures and implications**

Ye and Sugahara[1] describe a method of using multiple embeddings of multidimensional time series for building predictions of chaotic systems, in particular interconnected ecosystems. They use single nearest neighbor predictors from each of several different embeddings, averaging the k best (k=sqrt(n),n= number of embeddings), where best is determined by correlation of the predictions with observations in training.

For very high dimensional attractors, the number of such embeddings can quickly overcome computational capability (e.g. weather and climate). For example, choosing 12 dimensional embeddings using local temperature and rainfall plus 5 seasonally averaged empirical orthogonal functions (EOF's) such as the multivariate ENSOindex[13], the pacific decadal oscillation[14], the arctic oscillation[15], the North Atlantic oscillation[15], and the Atlantic multi decadal oscillation[16] over a 3 year time period results in over $10^{14}$ possible embeddings to search through. One approach is to then use a random sample of embeddings from this set. Here a theory is proposed for improving performance and reducing computation relative to using such a random sample of these embeddings. The theory is developed in the context of nearest neighbor averages[3,4], but applied to predictive distributions built from linear models and neural net models in a multi-season ahead precipitation prediction context and to a decision function based on whether a stock index will decrease or increase the next day.

The first result is a corollary of Sauer et al.'s[17] Theorems 2.7 and 2.10. The precise statement of the result requires a few mathematical definitions[17]. The box dimension of a compact set A in n dimensional Euclidean space is $box \dim(A) = \lim_{\varepsilon \to 0} \left( \frac{\log(N_\varepsilon)}{-\log(\varepsilon)} \right)$ where $N_\varepsilon$ is the number of cubes with side $\varepsilon$ that it takes to cover A. Again following Sauer (ibid) a general delay coordinate map takes the form $F(x) = \left( h_1(x), \ldots, h_1(g^{n_1-1}(x)), \ldots, h_j(x), \ldots, h_j(g^{n_j-1}(x)) \right)$, where $h_i(x)$ is the ith coordinate of a measurement of a term on a chaotic attractor, and $g^k(x)$ is k fold application of a diffeomorphism or flow on A. Let $n = n_1 + \ldots + n_j$. Theorem 2.7 of Sauer et al now states, given some restrictions on periodic orbits, that if $n > 2 box \dim(A)$ then F is both 1-1 on A and an *immersion* (derivatives are 1-1) on every smooth manifold C contained in A. This is the primary result motivating most embedding procedures.

One more definition is required for stating theorem 2.10 (ibid). The $\delta$-distant self intersection set is defined as $\Sigma(F, \delta) = \{ x \in A : F(x) = F(y) \text{ for some } y \in A, |x - y| > \delta \}$. Then theorem 2.10 states that if A is compact in $R^k$ with $box \dim(A) = d$, then if $n \leq 2d$, $\Sigma(F, D)$ has lower box counting dimension at most $2d - n$, and $F$ is an immersion on each compact subset C of an m-manifold contained in A except on a subset of C of dimension at most 2m-n-1.

Now define $F_1, F_2$ to be strictly distinct if the values of their respective $h_i$ do not overlap. The first result of this paper is then:

**Corollary 1:** For almost all $F_1, F_2$ strictly distinct with dimensions $n_1, n_2$, if $d < n_1 < n_2 \leq 2d$, then $\Sigma(F_1, \delta) \cap \Sigma(F_2, \delta) = \phi$.

PROOF: Suppose there is a set of $F_i, F_j$ strictly distinct, with positive probability such that $\phi \neq \Sigma(F_i, \delta) \cap \Sigma(F_j, \delta)$, for $i, j$. Then it would be possible to construct a set of product maps $(F_i, F_j)$ with positive probability for which $\phi \neq \Sigma((F_i, F_j), \delta)$, even though the box dimension of the product map is $n^{i,j} = n_i + n_j > 2d$, where $n^{i,j}$ is the box dimension of $(F_i, F_j)$ contradicting theorem 2.7 of Sauer (ibid).

Note that if $F_i, F_j$ are distinct, but not strictly distinct, the outcome then depends on the overlap. If the number of overlapping coordinates is such the $n^{i,j} > 2d$ the result holds. If $n^{i,j} \leq 2d$, the null set may be common, but cannot be larger than $2d - n^{i,j}$. By recursion it becomes plain that a large sample of predictors built on distinct delay maps of dimension $> d$ will provide significantly better predictions than a predictor built on a single delay map, and may save significant computational load by reducing the dimension of each model.

Suppose we wish to model $x$, a term in $F_i, i = 1,2$ and suppose that we have a consistent estimator of $E(x|F_i)$ (e.g. nearest neighbor estimates are consistent[3] but single hidden layer neural net models can be as well[18]) Suppose $d < n_i < 2d$. Let $\Sigma_i = \Sigma(F_i, \delta), i = 1,2$. Then we can decompose the attractor into 4 mutually exclusive sets,

$S_1 = \{\Sigma_1 \cap \Sigma_2\}, S_2 = \{\Sigma_1 \cap \overline{\Sigma}_2\}, S_3 = \{\overline{\Sigma}_1 \cap \Sigma_2\}, S_4 = \{(\overline{\Sigma}_1 \cap \overline{\Sigma}_2)\}$ Define $\sigma_i^2(x) = \int_{Si}((x - E_i(x))^2) d\mu$ and

$E_i(x) = \int_{Si}((x)) d\mu$ where $\mu$ is the normalized SRB[19] measure on the attractor. Suppose we use our hypothetical consistent estimator on each $F_i$ and look at the pairs of prediction based on each.

Exhibit 1

$E_4((E(x|F_1), E(x|F_2))) = (x, x)$

$E_3(((E(x|F_1), E(x|F_2)))) = (x, E_3(x, F_2))$

$E_2((E(x|F_1), E(x|F_2))) = (E_2(x|F_1), x)$

$E_1(E(x|F_1), E(x|F_2)) = (E_1(x|F_1), E_1(x|F_2))$

So the true value shows up in regions $S_2, S_3, S_4$. Taking advantage of corollary 1, making $F_1$, and $F_2$ strictly distinct, $S_1 = \phi$ so the true value of the attractor is at least one of the predictions.

To turn estimation into prediction with lead L, we need only insure that $x$, the variable to be predicted is the term in the delay map furthest forward in time, and it is L time steps ahead of the next term back. To lay to rest all difficulties about distinctness, we can consider $x$ appended to the delay map discussed in the theorems, with lead L over all the terms in the original map.

With a natural high dimensional system like climate, where we do not know $d$, we need to have a way of approaching the problem when n is much smaller than d. To allow some formal reasoning that results in testable hypothesis, consider the following definition of weak shadowing.

**Weak shadowing:** Suppose $F_{i,k}$ is a k dimensional sub delay map of delay map $F_i^*$ where $d < box\dim(F_i^*) < 2d$.

Then $F_{i,k}$ weakly shadows $F_i^*$ if outside of $\Sigma_i$: $\left( \int_{\Omega \backslash \Sigma_i} \left( \left( E(x|F_{i,k}) - E(x|F_i^*) \right)^2 \right) d\mu \right) < \left( \int_{\Omega \backslash \Sigma_i} \left( \left( x - \frac{\int_{\Omega \backslash \Sigma_i} x d\mu}{\int_{\Omega \backslash \Sigma_i} d\mu} \right)^2 \right) d\mu \right)$. (2)

In particular $F_{i,k}$ inherits some predictive power from $F_i^*$. Now we can state a conjecture:

**Conjecture 1:** *Under consistent statistical model selection procedures, assuming weak shadowing exists, disjoint sampling of delay maps provides a more efficient approach to multiview embedding and prediction then random sampling.*

The importance of the conjecture is that there exists a set of delay maps with very small dimension that give non trivial prediction at lead L, and distinct random sampling (without replacement sampling) is a more efficient way of finding them than pure random sampling. To test the conjecture in the next section, we will examine 6th season ahead prediction over 12 weather stations in the New York, New Jersey, Connecticut area. We will use 6[th] season ahead predictions made combining delay maps constructed from 10 random disjoint partitionings of the delay map space (120 total delay maps) and compare that to 6[th] season ahead predictions made from 1000 totally random delay maps in the tri state area.

The 1st conjecture focuses on the way we construct multiview embeddings to best complement one another. This conjecture is useful both in closed and open high dimensional system. The remaining development of this section focuses on open high dimensional dynamic systems, where the outside world can have an impact on the dynamics totally unrelated to the evolving dynamics of the system.

Linear response theory as applied to systems far from equilibrium[5] says that the linear models built from the natural fluctuations occurring in the unstable manifolds of the attractor predict the response of those manifolds to small disturbances. The problem is that typically high dimensional chaotic systems also have stable manifold which respond to disturbances by a decaying oscillation. The responses of the stable manifold will only occur in the presence of a disturbance, while the response of the unstable manifold will continuously be reflected in the evolution of the system. If the system is high enough dimension and the stable manifolds and responses are high enough dimension, then one can hope the response of the stable manifolds will tend toward a gaussian noise term[20], while the response of the unstable manifolds will be consistently represented. This leads to a second definition and conjecture.

Define an open high dimensional attractor and its environment to have a tenuous joint distribution of outside impulses and stable manifolds with respect to a measurable characteristic of the attractor if the effect of outside impulses on the stable manifolds is indistinguishable from adding exponentially decaying Gaussian noise to the measurable characteristic. Define a predictive decision as an action to be taken based on the difference between some threshold and the prediction. Then we can state the conjecture:

Conjecture 2: *When an attractor and its environment have a tenuous joint distribution of outside impulses and stable manifolds with respect to a measurable characteristic of the attractor, then linear response models on delay maps constructed from the history using statistical linear regression theory for independent and identically distributed errors will be predictive and will offer better predictive decisions than nearest neighbor models.*

When a closed system simulating the dynamic system exists (for example, a digital climate model) linear models built on multiview embeddings will reflect only the unstable manifold disturbances, so will predict quite well in the real world when far enough ahead so that the stable manifold oscillations have subsided. When the only source of data for model building is historical data then the stable manifold effect must be damped some other way, so that the consistent linear response to the unstable manifold is captured, and irrelevant fluctuations are ignored.

The latter approach is possible either when there is a natural threshold, and prediction beyond the threshold is important or when the stable manifold effect actually acts like white noise. One case where this can occur is when considering trying to time the stock market. If we simple put money in the stock market on the days the market will rise, and take it out on days that the market will fall, we can "beat" the market consistently and make more money than simply letting it sit in the market. In this case, whenever the difference at close between todays market and tomorrows market is negative, we take the money out. Otherwise we have the money in the market. Since the mean change in the market over very long times is positive, the threshold 0 is at a distance from the mean this fits the criteria of the conjecture.

Note in this case, with such a threshold, we are enriching the detection of fluctuations along the unstable manifolds over those in the stable manifold. It is also interesting to note that when non trivial lower dimensional linear response behavior exists, the chaotic system satisfies our definition of weak shadowing. In diffuse systems, longer periods of model building under stationary conditions should also result in improved prediction, as the noise effect will wash out of the coefficients.

### 3) Examples

Two examples are discussed in this section, 6th season ahead prediction of precipitation over 12 weather stations in the New York, New Jersey, Connecticut area, and day ahead predictions of the daily close of three Dow Jones market indexes.

For the precipitation prediction, we will use $6^{th}$ season ahead predictions made combining delay maps constructed from 10 random disjoint partitionings of the delay map space (120 total delay maps) and compare that to $6^{th}$ season ahead predictions made from 1000 totally random delay maps in the tri state area.

The predictions were constructed using linear models or approximate neural net models built on 100 year long runs of a climate model[8,9] at stable global surface air temperature[21]. The models were then applied to real data, and the correlations with a training set of real data with predictions were used to sort them, before testing them on a data further in the future.

In a preliminary estimate the decomposition approach not only reduced computation time by an order of magnitude, it appeared to be more accurate according to the following statistical test. In this test models are compared in a penultimate form, where a number of models have already been combined together, but they are still separated by:

1. Linear or neural net approximation based on linear models
2. Minimum correlation in the base line data accepted.
3. Basis seasons for calibration of the predictions[21], (8 seasons starting at the 1$^{st}$, 2$^{nd}$, 3$^{rd}$ or 4$^{th}$ after the final model selection data, predictions are all done beyond (further to the future than) this calibration data set, so are true predictions).

In addition within each data set there are either 1000 or 500 individual prediction sequences corresponding to randomly selected embeddings according to purely random selection (second case) or randomly partitioned distinct models( 1st case). The densities are produced by keeping separate predictions separated by:

A. Threshold correlation (large than that in 2 above)
B. Minimum Year index for the attractor (the year index is the year corresponding to the amount of green house gases assumed in the data, run to stabilized average global surface air temperature to get an attractor, this was done to allow an estimate of a stationary attractor to be used).
C. Proportion of the historic density mixed in with the purely "predictive" density (between .01 and .99, chosen randomly).

For each of the 500 or 1000 prediction sequences, each predicted season where the "predictive" density results in higher likelihood than the historic density is given a 1, otherwise it is a 0. A matrix of 1s and 0's is created with rows representing models and columns representing seasons.

The statistical test is designed to answer one particular question, "do the models at this penultimate stage show statistically significant predictive power over prediction built from purely random delay maps". Since determining what a randomly chosen density should look like to get a baseline comparison with the full raw data was hard to conceptualize, a conditional test is used.

For each season, condition on the models that predict well in that season, and then see if the conditional probability of predicting seasons further in the future is more likely for those models. If M is the matrix of 0's and 1's to estimate these conditional probabilities calculate $M^T M$. The diagonal is now the number of 1s in each row, and the rows of the upper triangular portion divided by the row diagonal is the conditional probability of correctly predicting given predicting correctly at the season represented by the diagonal (call the matrix corresponding to this CP). If N is the total number of rows of M, then diag($M^T M$)/N is the (horizontal) vector of base probabilities and P=(t(diag($M^T M$)/N)diag($M^T M$)/N) is the matrix of probabilities conditioning on the observed number of 1's in each column of M, and assuming they are independent. As a rough measure of distance between the P and the CP we can use (CP-P)/sqrt(P*(1-P)/N) (division here is element wise). Because of the intermittent prediction, a subset of the test statistic upper triangular matrix, the largest 4 elements are used for the calculation. To estimate the distribution of the test statistic under independence we permute the 1s and 0s in each column of M independently and recalculate the test statistic, a P value is based on the proportion of simulation test statistics greater than that observed out of 1000.

The results are shown in table 1 below. Clearly the distinct rows even though they required almost an order of magnitude less core time for calculation than the random, provide more information. Similar patterns hold whether the largest 2, 6, 16 or 20 or 70 deviations are included.

Table 1: P values of test statistics for a null hypothesis of 0 information in predictive densities comparing distinct vs purely random sampling of delay maps, $r_t$ is the minimum correlation allowed for delay map models from the

|  | Distinct lag 1 | Distinct lag 2 | Distinct lag 3 | Distinct lag 4 | Random lag 1 | Random lag 2 | Random lag 3 | Random lag 4 |
|---|---|---|---|---|---|---|---|---|
| Linear, $r_t=0.3$ | 0 | 0.988 | 0 | 0 | 1 | 0 | 0.816 | 0.075 |
| Linear, $r_t=0.0$ | 0 | 0.651 | 0 | 0 | 0.962 | 0 | 0.989 | 1 |
| Linear, $r_t=-1$ | 0 | 0.002 | 0 | 0 | 0.363 | 0.002 | 0.966 | 1 |
| NN, $r_t=0.3$ | 0.008 | 0.002 | 0 | 0.021 | 0.001 | 0.742 | 0.035 | 0.772 |
| NN, $r_t=0.0$ | 0 | 0 | 0 | 0 | 0.022 | 0.576 | 0.094 | 0.346 |
| NN, $r_t=-1$ | 0 | 0 | 0 | 0.023 | 0.042 | 0.014 | 0 | 0.246 |

Taking a Poisson upper bound for these probabilities, and using a false discovery rate[22] criterion on distinct and random models separately with a q value of 0.05 gives 5 out of 24 interesting results for random samples and 22 out of 24 interesting results for distinct samples. The distinct sampling method appears to improve prediction results over pure random sampling of delay maps.

The second example involves using predictive decision making in an automated trading algorithm for the stock market. An initial analysis was done on the S&P 500 from 2000-2007 to determine how far back I would sample the price history and how many regressor terms would be considered in each regression. The decision was to go back 40 days, and use 20 variables for each model (40,20). In the analysis below, the potential variable stock includes index pricing data going back 40 days, for all three indexes. The 8 year training period is used to predict the two years following the end of the 8 year training period, then the whole system is moved forward to cover the next two years etc. The algorithm averages 5 trading paths using different random embedding sequences.

Analyses below are done on the Dow Jones industrial average (currently 30 industrial stocks), Dow Jones Transportation average (currently 20 stocks), and Dow Jones Utility average (currently 15 stocks) from 1929 through 2006. In each case a total training period of 8 years is used, either 3, or 6 years are used to build the models and respectively 5, or 2 years are used to down select the models according to their prediction ability. From the 20 initial predictor variables, either ordinary least squares is used using all 20 variables or a combination of least angle regression, and least squares is used. For the least angle regression (lars) option, cross validation is used on the least angle regression to choose the best predictor variables[23]. The default settings in lars in R are used, and when no variables are chosen, a uniformly shrunken least squares estimate of the coefficients for the 20 variables are used.  Finally the top predictive models are chosen in the down selection stage, either the top 40%, or 80% are chosen. Figure 1 and table 2 show the results based on using lars to build predictions, while Figure 2 and table 3 show the results based on using least squares. The plots  in the figure show the multiplicitive increase in investment. The points show the change in value of an investment if is invested and kept in the market continuously. The lines show the change in value of an investment being traded out and in as the market is predicted to drop and rise. Figures 1 and 2 show this assuming 0 cost per trade. Figures 3 and 4 show the same respective plots assuming a cost of 3 basis points per trade. Tables 2-5 show respectively Sharpe's ratio calculated on a decadal basis for each trading procedure relative to the index it is following, as well as a sign test based on comparing the 2 year multiplicative increments provided by trading (based only on that two years of prediction) against the multiplicative increment provided by keeping money in the market.  Sharpe's ratio (typically calculated on an annual basis) takes the form of a t statistic telling how much better an investing procedure is than some base line, and is a standard financial measure of an investments performance. The reason for calculating on a decadal basis is that on an annual basis, the stock market is not necessarily a good base line. The sign test is more ad hoc and used as a simple sanity check because the market does not seem to reflect standard statistical theory.

Figure 1: Lars shrunken predictor estimates and assuming 0 cost for trades

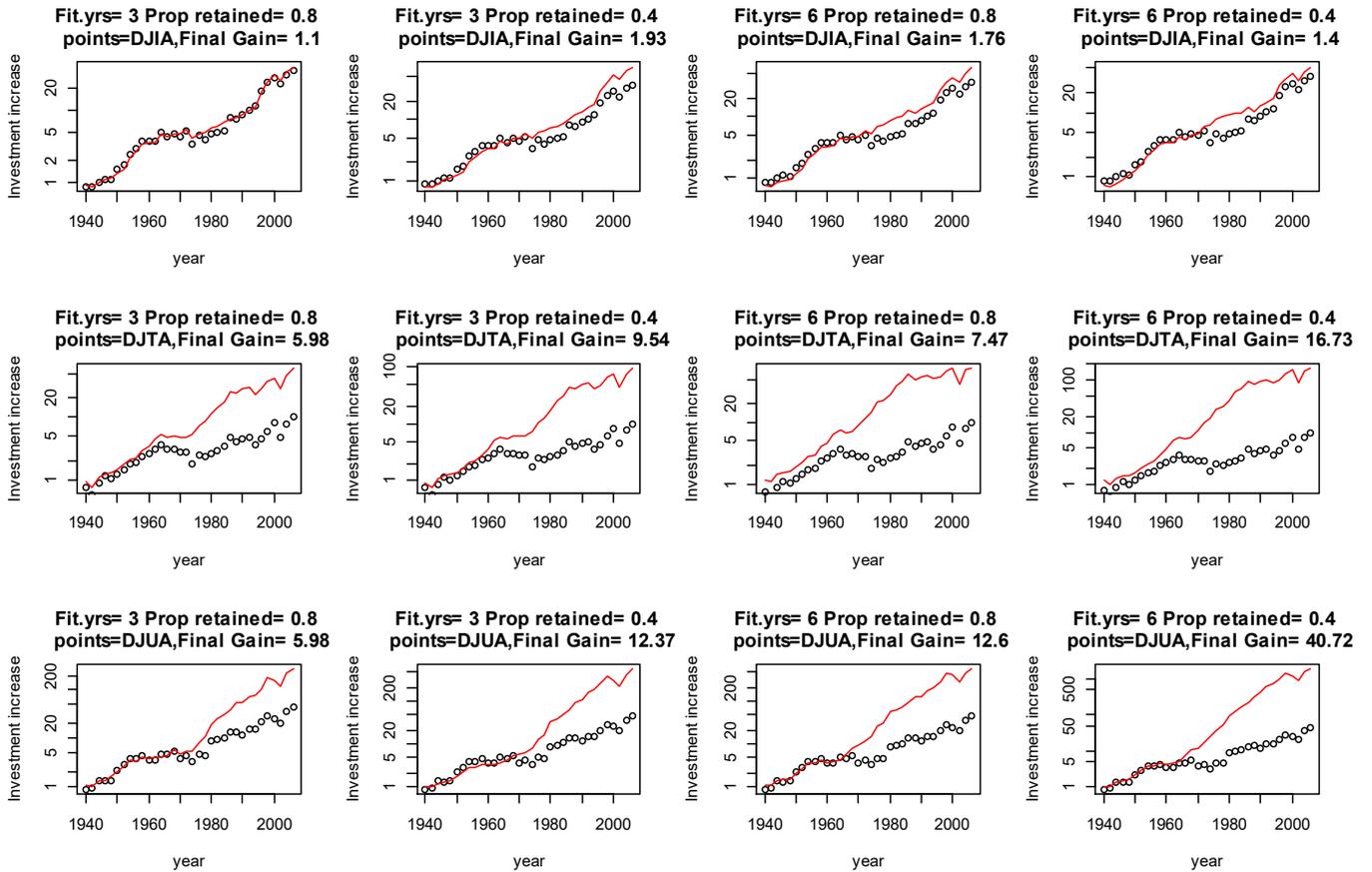

Figure 1: The columns keep the fit years and proportion retained constant, while the rows show the different Dow Indexes. Rough trends are increasing accuracy with increasing fit years and higher proportion of regressions rejected. Note the Investment increase is plotted on a log scale. Gain is how much the trading procedure changes the accumulated capital divided by how much the index does.

Table 2

| Sharpes Ratio/Sign test P value | 3 years Proportion retained 0.8 | 3 years Proportion retained 0.4 | 6 years Proportion retained 0.8 | 6 years Proportion retained 0.4 |
|---|---|---|---|---|
| DJIA | -.19/0.30 | 0.87/0/30 | 0.82/0.20 | 0.27/0.43 |
| DJTA | 2.20 /0.012* | 2.64/0.001* | 1.73/0.03* | 2.48/0.03* |
| DJUA | 1.76/0.06* | 2.26/0.06* | 2.27/0.06* | 3.13/0.012* |

*interesting at FDR of 0.1 under independence 0.3 without

Figure 2 Pure Least Squares prediction assuming 0 cost per trade

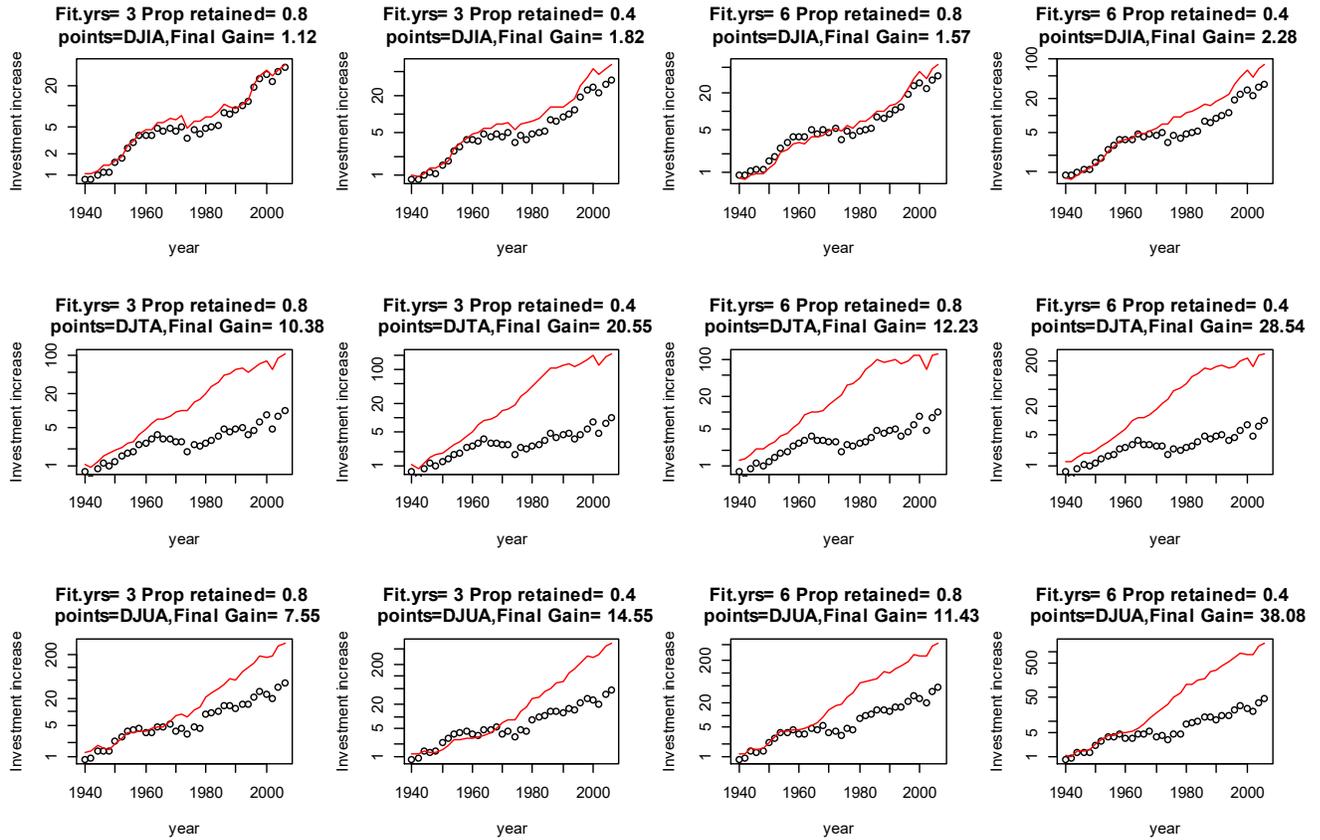

Figure 2: The columns keep the fit years and proportion retained constant, while the rows show the different Dow Indexes. Rough trends are increasing accuracy with increasing fit years and higher proportion of regressions rejected. Note the Investment increase is plotted on a log scale. Gain is how much the trading procedure changes the accumulated capital divided by how much the index does.

Table 3

| Sharpes Ratio/ Sign test P value | 3 years Proportion retained 0.8 | 3 years Proportion retained 0.4 | 6 years Proportion retained 0.8 | 6 years Proportion retained 0.4 |
|---|---|---|---|---|
| DJIA | -0.07/0.57 | 0.64/0.11 | 0.67/0.57 | 1.04/0.30 |
| DJTA | 2.35/0.06 | 2.92/0.01 | 2.12/0.11 | 2.89/0.06 |
| DJUA | 1.39/0.06 | 1.79/0/19 | 1.71/0.30 | 2.77/0.06 |

Figure 3 Lars shrunken regression estimates assuming 3 basis points per trade

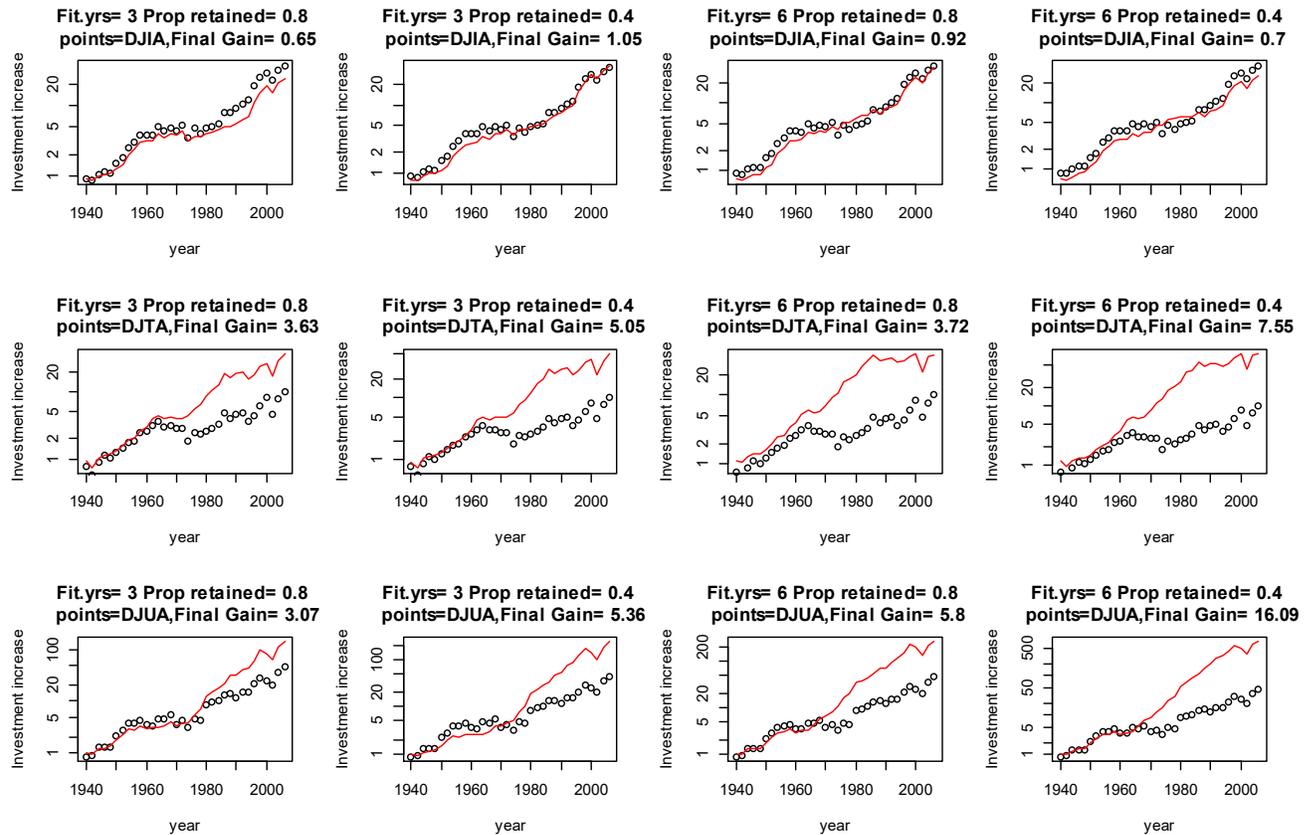

Figure 3: The columns keep the fit years and proportion retained constant, while the rows show the different Dow Indexes. Rough trends are increasing accuracy with increasing fit years and higher proportion of regressions rejected. Note the Investment increase is plotted on a log scale. Gain is how much the trading procedure changes the accumulated capital divided by how much the index does.

Table 4

| Sharpes Ratio/ Sign test P value | 3 years Proportion retained 0.8 | 3 years Proportion retained 0.4 | 6 years Proportion retained 0.8 | 6 years Proportion retained 0.4 |
|---|---|---|---|---|
| DJIA | -1.14/0.94 | -0.27/0.70 | -.51/0.70 | -.82/0.70 |
| DJTA | 1.47/0.06 | 1.83/0.005* | 1.01/0.20 | 1.70/0.11 |
| DJUA | 0.95/0.30 | 1.39/0.11 | 1.45/0.30 | 2.25/0.19 |

*interesting at FDR 0.1 assuming independence, 0.2 without.

Figure 4 Pure Least Squares predictions assuming 3 basis points per trade

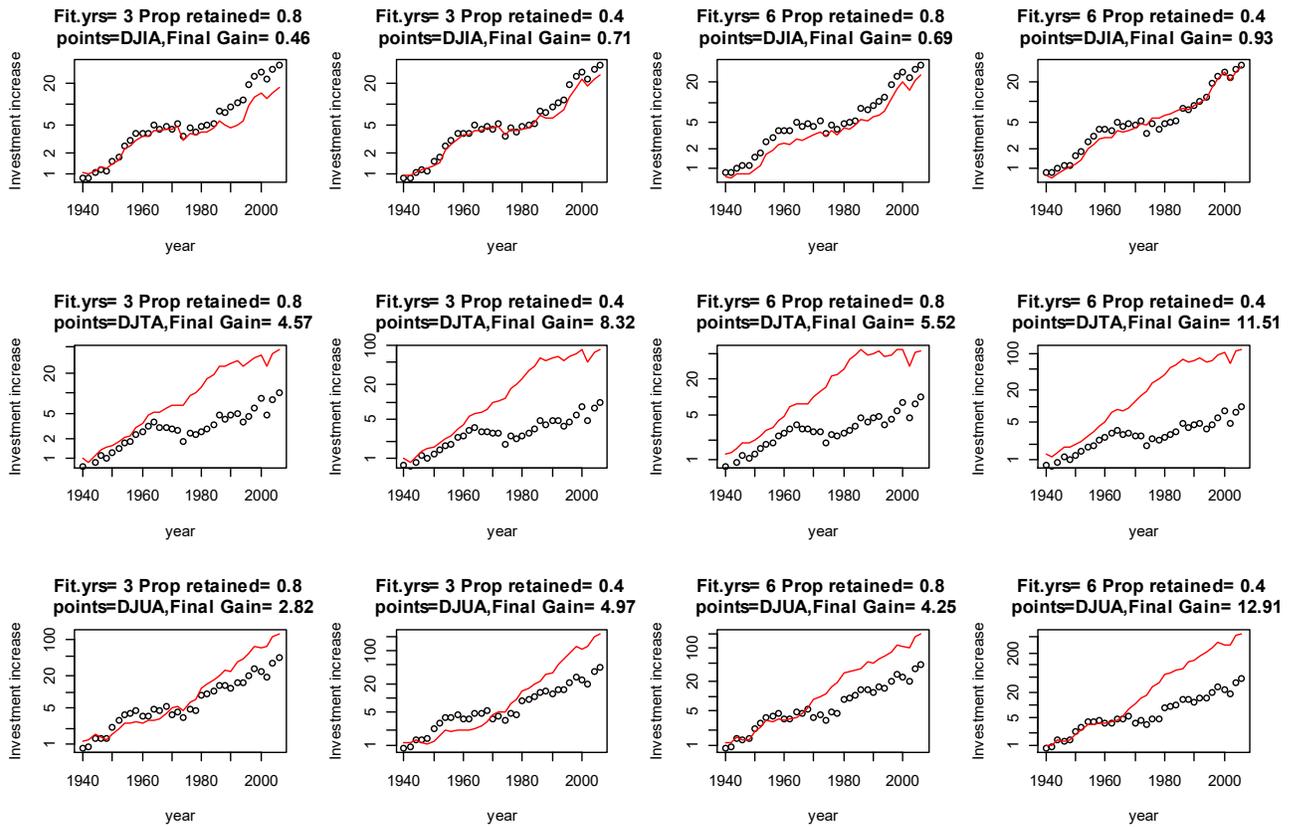

Figure 4: The columns keep the fit years and proportion retained constant, while the rows show the different Dow Indexes. Rough trends are increasing accuracy with increasing fit years and higher proportion of regressions rejected. Note the Investment increase is plotted on a log scale. Gain is how much the trading procedure changes the accumulated capital divided by how much the index does.

Table 5

| Sharpes Ratio/ Sign test P value | 3 years Proportion retained 0.8 | 3 years Proportion retained 0.4 | 6 years Proportion retained 0.8 | 6 years Proportion retained 0.4 |
|---|---|---|---|---|
| DJIA | -.17/0.89 | -.91/0.43 | -.91/0.80 | -.46/0.80 |
| DJTA | 1.26/0.20 | 1.88/0.11 | 1.35/0.11 | 2.0/0.11 |
| DJUA | 0.44/0.30 | 0.87/0.30 | 0.81/0.57 | 1.77/0.30 |

In all cases, what is being predicted is the change in the index from day to day (what is known in time series jargon as the first difference of the index data) the regressions are reduced either to the top 40% or 80% of predictors, and the prediction is a based on a timewise trimmed mean of those best predictors[19,21]. Probably the most interesting point is that the Pearson correlation between these predictors and the first differenced data is not overly impressive , because the magnitude of some of the deviations is so large, even though the correlation is statistically significant (Figure 5). Large magnitude deviations appearing on an unpredictable bases are expected from the chaos theory. However the direction and timing of the zero crossing appear to be predicted well enough to indicate that the stock market is to a certain extent predictable.

Figure 5 Daily predictions vs actual increments for all three indexes, Least squares top row, Lars fit bottom row

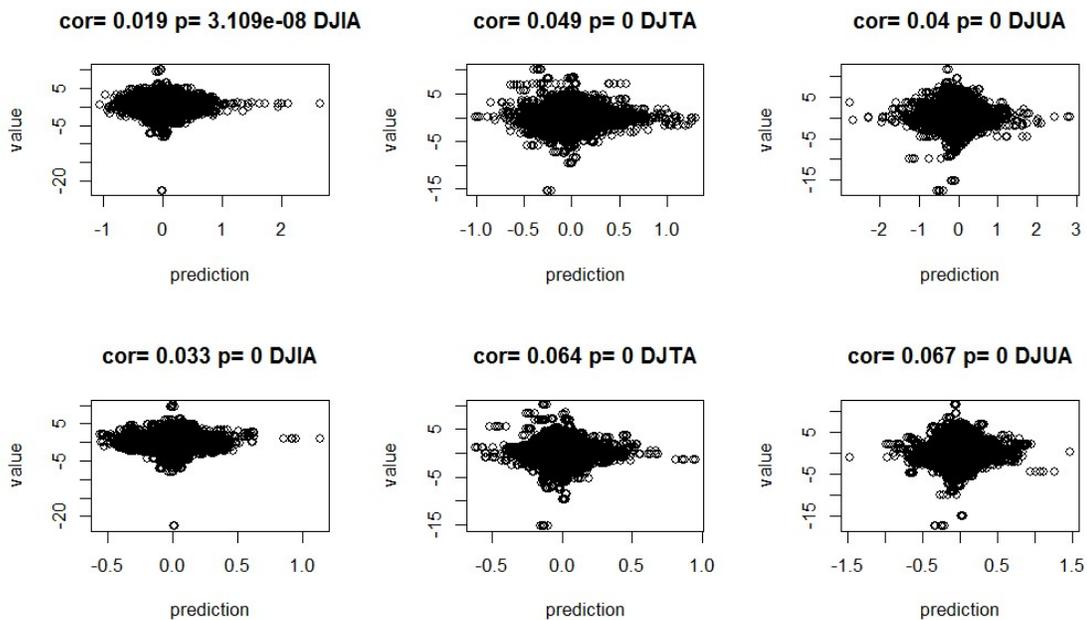

The statistical significance of the sign test and the correlations with the data are larger for the shrunken estimates, so the use of shrunken estimates should provide some advantage, however with respect to actual operation of the algorithm to increase money, (the gain in figures 1-4) the results are mixed. In summary, a simple modification of regression methodology based on embedology theory seems to allow quite good discrimination in the stock market when there is no mechanism to separate stable and unstable manifold effects. When there is, such as when a computer model is available, when not subject to significant external disturbances for fitting, then straight forward prediction over significant time periods seems possible and is improved by disjoint sampling.

With respect to the stock market predictions, there are two salient points. While this method does allow one to predict the stock market better than chance, even accounting for cost of trading, it does not beat the buy and hold strategy if yearly capital gains taxes are taken out. Hedge funds, which use a significantly larger base of possible predictors probably can with out to much problem (After all, the calculations were with freely available regression software using methods taught in typical advanced regression classes). In fact, a large number are probably profiting by these relationships. In as much as the chaos is driven by the actual values of the funds vs their current value this is well within the purpose of the stock market, but in the simple approach outlined here, where money is taken out of the market, when it will drop, and put back in when it will rise, it serves as an amplifier to any dynamic which the method can detect, thereby increasing the volatility in the market. So any such method must also be examined, not just from the point of view of its profitability, but also as to how it effects the primary purpose of the market, investment in increased productivity.